\documentclass[11pt]{article}

\usepackage[T1]{fontenc}              % T1 font encoding as default
\usepackage[latin9]{inputenc}         % ISO-8859-15 (Latin 9) text

\usepackage[a4paper]{geometry}        % DIN A4 paper

\usepackage{hyperref}
\usepackage{url}

\usepackage{booktabs}

\usepackage{graphicx} % WE USE PDFLATEX
\usepackage{amssymb,amsfonts,amsmath}
\usepackage{amsthm}

\usepackage{natbib}
\usepackage{mathpazo}
\usepackage{microtype}

\usepackage{color}

\usepackage{todonotes}
\usepackage{afterpage}
 
\usetikzlibrary{matrix}

\newcommand\bA{\boldsymbol A}
\newcommand\bB{\boldsymbol B}

\newcommand\bG{\boldsymbol G}

\newcommand\bI{\boldsymbol I}
\newcommand\bK{\boldsymbol K}

\newcommand\bM{\boldsymbol M}
\newcommand\bN{\boldsymbol N}
\newcommand\bQ{\boldsymbol Q}
\newcommand\bR{\boldsymbol R}
\newcommand\bS{\boldsymbol S}
\newcommand\bU{\boldsymbol U}
\newcommand\bV{\boldsymbol V}
\newcommand\bW{\boldsymbol W}
\newcommand\bX{\boldsymbol X}
\newcommand\bx{\boldsymbol x}
\newcommand\bY{\boldsymbol Y}
\newcommand\by{\boldsymbol y}
\newcommand\bZ{\boldsymbol Z}

\newcommand\bb{\boldsymbol b}
\newcommand\ba{\boldsymbol a}

\newcommand\btX{\widetilde{\boldsymbol X}}
\newcommand\btY{\widetilde{\boldsymbol Y}}

\newcommand\tX{\widetilde{X}}
\newcommand\tY{\widetilde{Y}}

\newcommand\bLambda{\boldsymbol \varLambda}
\newcommand\bmu{\boldsymbol \mu}
\newcommand\bPhi{\boldsymbol \varPhi}
\newcommand\bPsi{\boldsymbol \varPsi}
\newcommand\bRho{\boldsymbol P}
\newcommand\bSigma{\boldsymbol \varSigma}
\newcommand\bTheta{\boldsymbol \varTheta}

\newcommand\balpha{\boldsymbol \alpha}
\newcommand\bbeta{\boldsymbol \beta}

\newcommand\cor{\text{cor}}
\newcommand\cov{\text{cov}}
\newcommand\diag{\text{diag}}
\newcommand\expect{\text{E}}

\newcommand\trace{\text{Tr}}
\newcommand\var{\text{var}}
\newcommand\sgn{\text{sign}}

\newcommand\MSE{\text{MSE}}

\newcommand{\figcite}[1]{Fig.~\textbf{\ref{#1}}}
\newcommand{\eqcite}[1]{Eq.~\textbf{\ref{#1}}}

\newcommand{\parttitle}[1]{\noindent\underline{#1:}\ }

\begin{document}

\date{12 July 2018; revised 29 November 2018\\To appear in BMC Bioinformatics}

\title{
A Whitening Approach to Probabilistic Canonical Correlation Analysis for Omics Data Integration
}

\newcommand\myaddress[2][]{\relax
    {\noindent{\small$^{#1}$#2}}\\}

\author{Takoua Jendoubi$^{1,2}$\thanks{Email: {\tt t.jendoubi14@imperial.ac.uk} }
~\ and Korbinian Strimmer$^{3}$ \thanks{Email: {\tt korbinian.strimmer@manchester.ac.uk} }
}

\maketitle

\myaddress[1]{Epidemiology and Biostatistics, School of Public Health,
  Imperial College London, Norfolk Place, London W2 1PG, UK.}
\myaddress[2]{Statistics Section, Department of Mathematics, 
  Imperial College London, South Kensington Campus, London SW7 2AZ, UK.}
\myaddress[3]{School of Mathematics, University of Manchester, 
 Alan Turing Building, Oxford Road, Manchester M13 9PL, UK.}

\newpage

\begin{abstract} 
\parttitle{Background}
Canonical correlation analysis (CCA) is a classic statistical tool for investigating complex multivariate data. Correspondingly, it has found many diverse applications, ranging from molecular biology and medicine to social science and finance. Intriguingly, despite the importance and pervasiveness of CCA, only recently a probabilistic understanding of CCA is developing, moving from an algorithmic to a model-based perspective and enabling its application to large-scale settings.

\parttitle{Results}
Here, we revisit CCA from the perspective of statistical whitening of random variables and propose a simple yet flexible probabilistic model for CCA in the form of a two-layer latent variable generative model.  The advantages of this variant of probabilistic CCA include non-ambiguity of the latent variables,  provisions for negative canonical correlations, possibility of non-normal generative variables, as well as ease of interpretation on all levels of the model. In addition, we show that it lends itself to computationally efficient estimation in  high-dimensional settings using regularized inference. We test our approach to CCA analysis in simulations and apply it to two omics data sets illustrating the integration of gene expression data, lipid concentrations and methylation levels. 

\parttitle{Conclusions}
Our whitening approach to CCA provides a unifying perspective on CCA, linking together sphering procedures, multivariate regression and corresponding 
probabilistic generative models.  Furthermore, we offer an efficient computer implementation  in the ``whitening'' R package available at \url{https://CRAN.R-project.org/package=whitening}.
\end{abstract}

\newpage

\section*{Background}

Canonical correlation analysis (CCA) is a classic and highly versatile statistical approach to investigate the linear relationship between two sets of variables \citep{Hotelling1936,HardleSimar2015Chap16}.  CCA helps to decode complex dependency structures in multivariate data and to identify groups of interacting variables.  Consequently, it has numerous practical applications in molecular biology, for example omics data integration \citep {CaoLiu+2015} and network analysis \citep{HongChen+2013}, but also in many other areas such as econometrics or social science.

In its original formulation CCA is viewed as an algorithmic procedure optimizing a set of objective functions, rather than as a probablistic model for the data. Only relatively recently this perspective has changed. \citet{BachJordan2005} proposed a latent variable model for CCA building on earlier work on probabilistic principal component analysis (PCA) by \citet{TippingBishop1999}. The probabilistic approach to CCA not only allows to derive the classic CCA algorithm  but also  provide an avenue for Bayesian variants \citep{Wang2007,KlamiKaski2007}. 

In parallel to establishing probabilistic CCA  the classic CCA approach has also been further developed in the last decade by introducing variants of the CCA algorithm that are more pertinent for high-dimensional data sets now routinely collected in the life and physical sciences. In particular, the problem of singularity  in the original CCA algorithm is resolved by introducing sparsity and regularization \citep{Waaijenborg+2008,Parkhomenko+2009,WittenTibshiraniHastie2009,HardoonShaweTaylor2011,WilmsCroux2015} and, similarly, large-scale computation is addressed by new algorithms \citep{CruzCanoLee2014,MaLuFoster2015}.

In this note, we revisit both classic and probabilistic CCA from the perspective of whitening of random variables \citep{KessyLewinStrimmer2018}.  As a result, we propose a simple yet flexible probabilistic model for CCA linking together multivariate regression, latent variable models, and high-dimensional estimation.  Crucially, this model for CCA not only facilitates  comprehensive understanding of both classic and probabilistic CCA via the process of whitening but also extends CCA by allowing for negative canonical correlations and providing the flexibility to include non-normal latent variables.

The remainder of this paper is as follows.  First, we present our main results. After reviewing classical CCA we demonstrate that the classic CCA algorithm is special form of whitening. Next, we show that  the link of CCA with multivariate regression leads to a probabilistic two-level latent variable model for CCA that directly reproduces classic CCA without any rotational ambiguity. Subsequently,  we discuss our approach by applying it to both synthetic data as well as to multiple integrated omics data sets. Finally, we describe our implementation in R and highlight computational and algorithmic aspects.

Much of our discussion is framed in terms of random vectors and their properties rather than in terms of data matrices. This allows us to study the probabilistic model underlying CCA separate from associated statistical procedures for estimation. 

\subsection*{Multivariate notation}

We consider two random vectors $\bX =( X_1, \ldots, X_p)^T$ and $\bY= (Y_1, \ldots, Y_q)^T$ of dimension $p$ and $q$. Their respective multivariate distributions $F_{\bX}$ and $F_{\bY}$ have expectation $\expect(\bX) = \bmu_{\bX}$ and $\expect(\bY) = \bmu_{\bY}$ and covariance $\var(\bX) = \bSigma_{\bX}$ and $\var(\bY) = \bSigma_{\bY}$. The cross-covariance between $\bX$ and $\bY$ is given by $\cov(\bX, \bY) = \bSigma_{\bX \bY}$. The corresponding correlation matrices are denoted by $\bRho_{\bX}$, $\bRho_{\bY}$, and $\bRho_{\bX \bY}$. By $\bV_{\bX} = \diag(\bSigma_{\bX}) $ and $\bV_{\bY}=\diag(\bSigma_{\bY})$ we refer to the diagonal matrices containing the variances only, allowing to decompose covariances as  $\bSigma = \bV^{1/2} \bRho \bV^{1/2}$. The composite vector $(\bX^T, \bY^T)^T$ has therefore mean $(\bmu_{\bX}^T, \bmu_{\bY}^T)^T$ and  covariance $\left( \begin{array}{c c} \bSigma_{\bX} & \bSigma_{\bX \bY}\\ \bSigma_{\bX \bY}^T & \bSigma_{\bY} \end{array}\right)$.

Vector-valued samples of the random vectors $\bX$ and $\bY$ are denoted by $\bx_i$ and $\by_i$ so that $ (\bx_1, \ldots, \bx_i, \ldots, \bx_n)^T$ is the $n \times p$ data matrix for $\bX$ containing $n$ observed samples (one in each row).  Correspondingly, the empirical mean for $\bX$ is given by  $\hat\bmu_{\bX} = \bar \bx =\frac{1}{n} \sum_{i=1}^n \bx_i$, the unbiased covariance estimate is $\widehat\bSigma_{\bX} =  \bS_{\bX} = \frac{1}{n-1} \sum^n_{i=1} (\bx_{i} -\bar{\bx})(\bx_{i} -\bar{\bx})^T$, and the corresponding correlation estimate is denoted by $\widehat\bRho_{\bX} = \bR_{\bX}$.

\section*{Results}

We first introduce CCA from a classical perspective, then we demonstrate that CCA is best understood as a special and uniquely defined type of whitening transformation. Next, we investigate the close link of CCA with multivariate regression. This not only allows to interpret CCA as regression model and to better understand canonical correlations,  but also provides the basis for a probabilistic generative latent variable model of CCA based on whitening. This model is introduced in the last subsection.

\subsection*{Classical CCA}

In canonical correlation analysis the aim is to find mutually orthogonal pairs of maximally correlated linear combinations of the components of $\bX$ and of $\bY$. Specifically, we seek canonical directions $\balpha_i$ and $\bbeta_j$ (i.e.\ vectors of dimension $p$ and $q$, respectively) for which~
\begin{equation}
\cor(\balpha_i^T \bX, \bbeta_j^T \bY)=
\begin{cases}
\lambda_i & \text{maximal for $i=j$}\\
0 & \text{otherwise, }
\end{cases}
\label{eq:ccaoptim1}
\end{equation}
where  $\lambda_i$ are the canonical correlations,
and simultaneously
\begin{equation}
\cor(\balpha_i^T \bX, \balpha_j^T \bX) =
\begin{cases}
1 & \text{for $i=j$}\\
0 & \text{otherwise, }
\end{cases}
\end{equation}
and 
\begin{equation}
\cor(\bbeta_i^T \bY, \bbeta_j^T \bY) = 
\begin{cases}
1 & \text{for $i=j$}\\
0 & \text{otherwise.}
\end{cases}
\end{equation}
In matrix notation, with  $\bA = (\balpha_1, \ldots, \balpha_p)^T$, $\bB =  (\bbeta_1, \ldots, \bbeta_q)^T$, and $\bLambda = \diag(\lambda_i)$, the above can be written as $\cor(\bA \bX, \bB \bY) = \bLambda$ as well as $\cor(\bA \bX) = \bI$ and $\cor(\bB \bY) = \bI$.   The projected vectors $\bA \bX$ and $\bB \bY$ are also called the CCA scores or the canonical variables.

\citet{Hotelling1936} showed that there are, assuming full rank covariance matrices $\bSigma_{\bX}$ and $\bSigma_{\bY}$, exactly $m = \min(p,q)$ canonical correlations and pairs of canonical directions $\balpha_i$ and $\bbeta_i$, and that these can be computed analytically from a generalized eigenvalue problem \citep[e.g.][]{HardleSimar2015Chap16}.   Further below we will see how canonical directions and correlations follow almost effortlessly from a whitening perspective of CCA. 

Since correlations are invariant against rescaling, optimizing \eqcite{eq:ccaoptim1} determines the canonical directions $\balpha_i$ and $\bbeta_i$ only up to their respective lengths, and we can thus arbitrarily fix the magnitude of the vectors $\balpha_i$ and $\bbeta_i$.  A common choice is to simply normalize them to unit length so that $\balpha_i^T \balpha_i= 1$ and  $\bbeta_i^T \bbeta_i= 1$.   

Similarly, the overall sign of the canonical directions $\balpha_i$ and $\bbeta_j$ is also undetermined.  As a result, different implementations of CCA may yield canonical directions with different signs, and depending on the adopted convention this can be used either to enforce positive or to allow negative canonical correlations, see below for further discussion in the light of CCA as a regression model.

Because it optimizes correlation, CCA is invariant against location translation of the original vectors $\bX$ and $\bY$, yielding identical canonical directions and correlations in this case.  However, under  scale transformation of $\bX$ and $\bY$ only the canonical correlations $\lambda_i$ remain invariant whereas the directions will differ as they depend on the variances $\bV_{\bX}$ and $\bV_{\bY}$. Therefore, to facilitate comparative analysis and interpretation the canonical directions the random vectors $\bX$ and $\bY$ (and associated data) are often standardized.

Classical CCA uses the empirical covariance matrix $\bS$ to obtain canonical correlations and directions.  However, $\bS$ can only be safely employed if the number of observations is much larger than the dimensions of either of the two random vectors $\bX$ and $\bY$, since otherwise $\bS$ constitutes only a poor estimate of the underlying covariance structure and in addition may also become singular. Therefore, to render CCA applicable to small sample high-dimensional data two main strategies are common: one is to directly employ regularization on the level of the covariance and correlation matrices to stabilize and improve their estimation; the other is to devise probabilistic models for CCA to facilitate application of Bayesian inference and other regularized statistical procedures.

\subsection*{Whitening transformations and CCA}

\subsubsection*{Background on whitening}

Whitening, or sphering, is a linear statistical transformation that converts a random vector $\bX$ with covariance matrix $\bSigma_{\bX}$ into a random vector 
\begin{equation}
\btX = \bW_{\bX} \bX
\label{eq:whitening}
\end{equation}
with unit diagonal covariance  $\var( \btX ) =\bSigma_{\btX} = \bI_p$.  The matrix $\bW_{\bX}$ is called the \emph{whitening matrix} or \emph{sphering matrix} for $\bX$,  also known as the \emph{unmixing matrix}.  In order to achieve whitening the matrix $\bW_{\bX}$ has to satisfy the condition $\bW_{\bX} \bSigma_{\bX} \bW_{\bX}^T = \bI_p$, but this by itself is not sufficient to completely identify $\bW_{\bX}$. There are still infinitely many possible whitening transformations, and the family of whitening matrices for $\bX$ can be written as 
\begin{equation}
\bW_{\bX} = \bQ_{\bX}  \bRho_{\bX}^{-1/2} \bV_{\bX}^{-1/2}\,.
\label{eq:whitencor}
\end{equation}
Here, $\bQ_{\bX}$ is an orthogonal matrix; therefore the whitening matrix $\bW_{\bX}$ itself is not orthogonal unless $\bRho_{\bX} =  \bV_{\bX} = \bI_p$.  The choice of $\bQ_{\bX}$ determines the type of whitening \citep{KessyLewinStrimmer2018}. For example, using $\bQ_{\bX} = \bI_p$ leads to ZCA-cor whitening, also known as Mahalanobis whitening based on the correlation matrix. PCA-cor whitening, another widely used sphering technique, is obtained by setting $\bQ_{\bX} = \bG^T$, where $\bG$ is the eigensystem resulting from the spectral decomposition of the correlation matrix $\bRho_{\bX} =\bG \bTheta \bG^T$.  Since there is a sign ambiguity in the eigenvectors $\bG$ we adopt the convention of \citet{KessyLewinStrimmer2018} to adjust columns signs of $\bG$, or equivalently row signs of $\bQ_{\bX}$,  so that the rotation matrix $\bQ_{\bX}$ has a positive diagonal.

The corresponding inverse relation $\bX = \bW_{\bX}^{-1} \btX =\bPhi_{\bX}^T  \btX$ is called a \emph{coloring} transformation, where  the matrix  $\bW_{\bX}^{-1} =\bPhi_{\bX}^T $ is the \emph{mixing matrix}, or \emph{coloring matrix} that we can write in terms of rotation matrix $\bQ_{\bX}$ as
\begin{equation}
\bPhi_{\bX} = \bQ_{\bX} \bRho_{\bX}^{1/2} \bV_{\bX}^{1/2}
\end{equation}
Like $\bW_{\bX}$ the mixing matrix $\bPhi_{\bX}$ is not orthogonal. The entries of the matrix $\bPhi_{\bX}$ are called the \emph{loadings}, i.e.\ the coefficients linking the whitened variable $\btX$ with the original $\bX$. Since $\btX$ is a white random vector with $\cov(\btX) = \bI_p$ the loadings are equivalent to the covariance $\cov(\btX, \bX) = \bPhi_{\bX}$. The corresponding correlations, also known as \emph{correlation-loadings}, are 
\begin{equation}
\cor(\btX, \bX) = \bPsi_{\bX} = \bPhi_{\bX} \bV_{\bX}^{-1/2}=\bQ_{\bX} \bRho_{\bX}^{1/2}\,.
\end{equation}
Note that the sum of squared correlations in each column of $\bPsi_{\bX}$ sum up to 1, as  $\diag(\bPsi_{\bX}^T\bPsi_{\bX}) = \diag(\bRho_{\bX}) = \bI_p$.

\subsubsection*{CCA whitening}

We will show now that CCA has a very close relationship to whitening. In particular, the objective of CCA can be seen to be equivalent to simultaneous whitening of both $\bX$ and $\bY$, with a diagonality constraint on the cross-correlation matrix between the whitened $\btX$ and $\btY$.

First, we make the choice to standardize the canonical directions $\balpha_i$ and $\bbeta_i$ according to $\var(\balpha_i^T \bX) = \balpha_i^T \bSigma_{\bX} \balpha_i= 1$ and $\var(\bbeta_i^T \bY) = \bbeta_i^T \bSigma_{\bY} \bbeta_i= 1$. As a result $\balpha_i$ and $\bbeta_i$ form the basis of two whitening matrices, $\bW_{\bX} = (\balpha_1, \ldots, \balpha_p)^T = \bA$  and $\bW_{\bY} = (\bbeta_1, \ldots, \bbeta_q)^T = \bB$, with \emph{rows} containing the canonical directions. The length constraint $\balpha_i^T \bSigma_{\bX} \balpha_i= 1$ thus becomes $\bW_{\bX} \bSigma_{\bX}\bW_{\bX}^T = \bI_p$  meaning that $\bW_{\bX}$ (and $\bW_{\bY}$) is indeed a valid whitening matrix.

Second, after whitening $\bX$ and $\bY$ individually to $\btX$ and $\btY$ using $\bW_{\bX}$ and $\bW_{\bY}$, respectively, the joint covariance of $(\btX^T, \btY^T)^T$ is 
$\left( \begin{array}{c c} \bI_p & \bRho_{\btX \btY}\\
 \bRho_{\btX \btY}^T & \bI_q \end{array}\right)$. 
Note that whitening of $(\bX^T, \bY^T)^T$ simultaneously would in contrast lead to a fully diagonal covariance matrix. In the above $\bRho_{\btX \btY}  = \cor( \btX , \btY) =  \cov( \btX , \btY)$ is the cross-correlation matrix between the two whitened vectors and can be expressed as
\begin{equation}
\bRho_{\btX \btY} =  \bW_{\bX} \bSigma_{\bX \bY} \bW_{\bY}^T = 
\bQ_{\bX} \bK \bQ_{\bY}^T = (\widetilde{\rho}_{ij}) 
\end{equation}
and
\begin{equation}
\bK = \bRho_{\bX}^{-1/2} \bRho_{\bX \bY} \bRho_{\bY}^{-1/2} = (k_{ij}).
\label{eq:Kdef}
\end{equation}
Following the terminology in \citet{ZS2011} we may call $\bK$ the correlation-adjusted cross-correlation matrix between $\bX$ and $\bY$.

With this setup the CCA objective can be framed simply as the demand that  $\cor( \btX , \btY)= \bRho_{\btX \btY}$ must be diagonal. Since in whitening the orthogonal matrices $\bQ_{\bX}$ and $\bQ_{\bY}$ can be freely selected we can achieve diagonality of $\bRho_{\btX \btY}$ and hence pinpoint the CCA whitening matrices by applying singular value decomposition to
\begin{equation}
\bK= (\bQ_{\bX}^\text{CCA})^T \bLambda \bQ_{\bY}^\text{CCA} \,.
\label{eq:Kdecomp}
\end{equation}
This provides the rotation matrices $\bQ_{\bX}^\text{CCA} $ and the $\bQ_{\bY}^\text{CCA} $ of dimensions $m \times p$ and $m \times q$, respectively, and the $m \times m$ matrix $\bLambda = \diag(\lambda_i) $ containing the singular values of $\bK$, which are also the singular values of $\bRho_{\btX \btY}$.  Since $m = \min(p,q)$ the larger of the two rotation matrices will not be a square matrix but it can nonetheless be used for whitening via \eqcite{eq:whitening} and \eqcite{eq:whitencor} since it still is semi-orthogonal with $\bQ_{\bX}^\text{CCA} (\bQ_{\bX}^\text{CCA})^T = \bQ_{\bY}^\text{CCA} (\bQ_{\bY}^\text{CCA})^T = \bI_m$. As a result, we obtain $\cor( \tX^{\text{CCA}}_i , \tY^{\text{CCA}}_i ) = \lambda_i$ for $i = 1 \ldots m$, i.e.\ the canonical correlations are identical to the singular values of~$\bK$.

Hence, CCA may be viewed as the outcome of a uniquely determined whitening transformation with underlying sphering matrices $\bW_{\bX}^{\text{CCA}}$ and $\bW_{\bY}^{\text{CCA}}$ induced by the rotation matrices $\bQ_{\bX}^\text{CCA}$ and  $\bQ_{\bY}^\text{CCA}$. Thus, the distinctive feature of CCA whitening, in contrast to other common forms of whitening described in \citet{KessyLewinStrimmer2018}, is that by construction it is not only informed by  $\bRho_{\bX}$ and  $\bRho_{\bY}$ but also by $\bRho_{\bX \bY}$, which fixes all remaining rotational freedom.

\subsection*{CCA and multivariate regression}

\subsubsection*{Optimal linear multivariate predictor}

In multivariate regression the aim is to build a model that, given an input vector $\bX$, predicts a vector $\bY$ as well as possible according to a specific measure such as squared error. Assuming a linear relationship $\bY^{\star} = \ba + \bb^T \bX$ is the predictor random variable, with mean $\expect(\bY^{\star}) = \bmu_{\bY^{\star}} = \ba+\bb^T \bmu_{\bX}$. The expected squared difference between $\bY$ and $\bY^{\star}$, i.e.\ the mean squared prediction error 
\begin{equation}
\begin{split}
\MSE 
&= \trace ( \expect( (\bY - \bY^{\star})  (\bY - \bY^{\star})^T ) ) \\
&= \sum_{i=1}^q  \expect( (Y_i-Y_i^{\star} )^2) ,
\end{split}
\end{equation}
is a natural measure of how well $\bY^{\star}$ predicts $\bY$. As a function of the model parameters $\ba$ and $\bb$ the predictive MSE becomes
\begin{equation}
\begin{split}
\MSE(\ba, \bb)  = &\trace((\bmu_{\bY}-\bmu_{\bY^{\star}}) (\bmu_{\bY}-\bmu_{\bY^{\star}})^T +\\
& \bSigma_{\bY} + \bb^T \bSigma_{\bX} \bb -2 \bb^T \bSigma_{\bX \bY}
) \,.
\end{split}
\end{equation}
Optimal parameters for best linear predictor are found by minimizing this MSE function. For the offset $\ba$ this yields 
\begin{equation}
\ba^{.} = \bmu_{\bY} - (\bb^{.})^T  \bmu_{\bX}
\end{equation} which regardless of the value of 
$\bb^{.}$ ensures
 $\bmu_{\bY^{\star}} - \bmu_{\bY} = 0$. 
Likewise, for the matrix of regression coefficients minimization results in
\begin{equation}
\bb^{\text{all}} = \bSigma_{\bX}^{-1}\bSigma_{\bX \bY}
\end{equation} 
with minimum achieved $\MSE(\ba^{\text{all}}, \bb^{\text{all}}) = \trace (\bSigma_{\bY} ) - \trace (\bSigma_{\bY \bX} \bSigma_{\bX}^{-1} \bSigma_{\bX \bY} )$. 

If we exclude predictors from the model by setting regression coefficients $\bb^{\text{zero}}=0$ then the corresponding optimal intercept is $\ba^{\text{zero}} = \bmu_{\bY}$ and the minimum achieved  $\MSE(\ba^{\text{zero}}, \bb^{\text{zero}}) = \trace (\bSigma_{\bY} )$. Thus, by adding predictors $\bX$ to the model the predictive $\MSE$ is reduced, and hence the fit of the model correspondingly improved, by the amount
\begin{equation}
\begin{split}
\Delta &= \MSE(\ba^{\text{zero}}, \bb^{\text{zero}})-\MSE(\ba^{\text{all}}, \bb^{\text{all}}) \\
 &= \trace (\bSigma_{\bY \bX} \bSigma_{\bX}^{-1} \bSigma_{\bX \bY} ) \\
&= \trace( \cov(\bY, \bY^{\text{all}\star})) \,.
\end{split}
\end{equation}
If the response $Y$ is univariate ($q=1$) then $\Delta$ reduces to the variance-scaled coefficient of determination $\sigma^2_Y \bRho_{Y \bX} \bRho_{\bX}^{-1} \bRho_{\bX Y}$. Note that in the above no distributional assumptions are made other than specification of means and covariances.

\subsubsection*{Regression view of CCA}

The first step to understand CCA as a regression model is to consider multivariate regression between two whitened vectors $\btX$ and $\btY$ (considering whitening of any type, including but not limited to CCA-whitening).  Since $\bSigma_{\btX} = \bI_p$ and $\bSigma_{\btX \btY} = \bRho_{\btX \btY}$ the optimal regression coefficients to predict $\btY$ from $\btX$ are given by 
\begin{equation}
\bb^{\text{all}} = \bRho_{\btX \btY}\, ,
\end{equation}
i.e.\ the pairwise correlations between the elements of the two vectors  $\btX$ and $\btY$. Correspondingly, the decrease in predictive MSE due to including the predictors $\btX$ is 
\begin{equation}
\begin{split}
\Delta & = \trace(\bRho_{\btX \btY}^T \bRho_{\btX \btY}) = \sum_{i,j}  \widetilde{\rho}^2_{ij} \\
       &= \trace(\bK^T \bK) = \sum_{i,j} k_{ij}^2 \\
       & = \trace(\bLambda^2) =\sum_i \lambda_i^2 \, .
\end{split}
\label{eq:delta}
\end{equation}

In the special case of CCA-whitening the regression coefficients further simplify to  $b^{\text{all}}_{ii} = \lambda_i$, i.e.\ the canonical correlations $\lambda_i$ act as the regression coefficients linking CCA-whitened $\btY$ and $\btX$.  Furthermore, as the decrease in predictive MSE $\Delta$ is the sum of the squared canonical correlations (cf. \eqcite{eq:delta}), each $\lambda_i^2$  can be interpreted as the variable importance of the corresponding variable in $\btX^{\text{CCA}}$ to predict the outcome $\btY^{\text{CCA}}$.  Thus, CCA directly results from multivariate regression between CCA-whitened random vectors, where the canonical correlations $\lambda_i$ assume the role of regression coefficients and $\lambda_i^2$ provides a natural measure to rank the canonical components in order of their respective predictive capability.
 
A key difference between classical CCA  and regression is that in the latter both positive and negative coefficients are allowed to account for the directionality of the influence of the predictors. In contrast, in classical CCA only positive canonical correlations are permitted by convention.  To reflect that CCA analysis is inherently a regression model we advocate here that  canonical correlations should indeed be allowed to assume both positive and negative values, as fundamentally they are regression coefficients. This can be implemented by exploiting the  sign ambiguity in the singular value decomposition of $\bK$ (\eqcite{eq:Kdecomp}). In particular, the rows signs of $\bQ_{\bX}^\text{CCA}$ and $\bQ_{\bY}^\text{CCA}$ and the signs of $\lambda_i$ can be revised simultaneously without affecting  $\bK$. We propose to choose $\bQ_{\bX}^\text{CCA}$ and $\bQ_{\bY}^\text{CCA}$ such that both rotation matrices have a positive diagonal, and then to adjust the signs of the $\lambda_i$ accordingly. Note that orthogonal matrices with positive diagonals are closest to the identity matrix (e.g. in terms of the Frobenius norm) and thus constitute minimal rotations.

\subsection*{Generative latent variable model for CCA}

\begin{figure*}[t]
\begin{center}
\includegraphics[scale=.8]{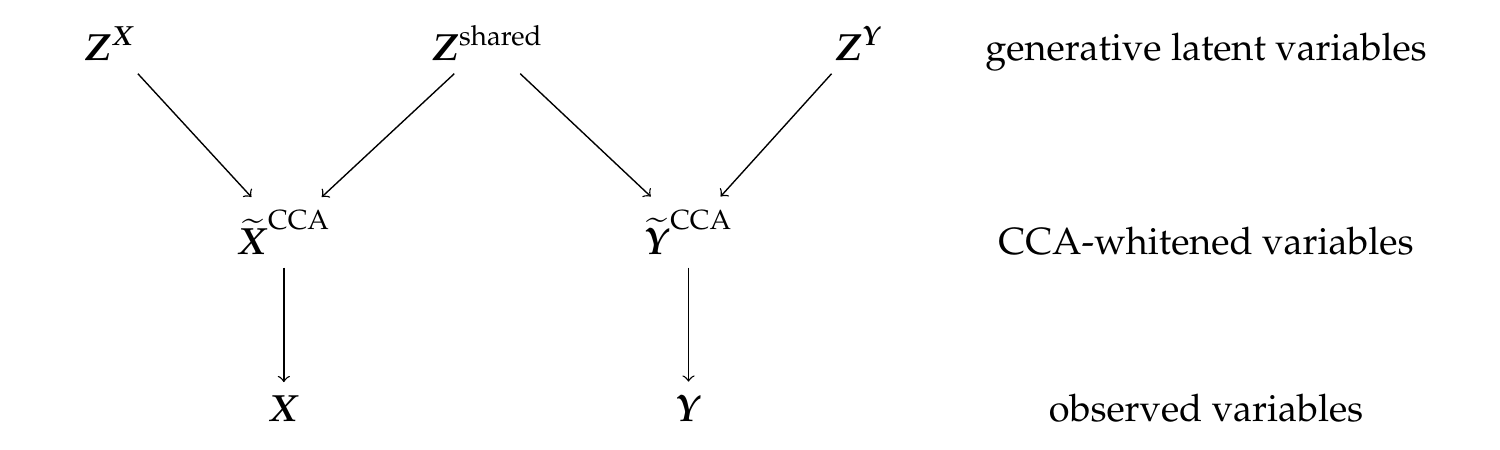}
\caption{Probabilistic CCA as a two layer latent variable generative model. The middle layer contains the CCA-whitened variables $\btX^{\text{CCA}}$ and $\btY^{\text{CCA}}$, and the top layer the uncorrelated generative latent variables  $\bZ^{\bX}$, $\bZ^{\bY}$, and $\bZ^{\text{shared}}$. \label{fig:probcca}} 
\end{center}
\end{figure*}

With the link of CCA to whitening and multivariate regression established it is straightforward to arrive at simple and easily interpretable generative probabilistic latent variable model for CCA. This model has two levels of hidden variables: it uses uncorrelated latent variables $\bZ^{\bX}$, $\bZ^{\bY}$, $\bZ^{\text{shared}}$ (level~1) with zero mean and unit variance to generate the CCA-whitened variables $\btX^{\text{CCA}}$ and $\btY^{\text{CCA}}$ (level~2) which in turn produce the observed vectors $\bX$ and $\bY$ -- see \figcite{fig:probcca} 

Specifically, on the first level we have latent variables
\begin{equation}
\begin{split}
\bZ^{\bX} &\sim F_{\bZ_{\bX}}, \\
\bZ^{\bY} &\sim F_{\bZ_{\bY}}, \,\text{and}\\
 \bZ^{\text{shared}} &\sim F_{\bZ_{\text{shared}}},
\end{split}
\end{equation}
with $\expect(\bZ^{\bX}) = \expect(\bZ^{\bY}) = \expect(\bZ^{\text{shared}}) = 0$ and $\var(\bZ^{\bX}) = \bI_p$, $\var(\bZ^{\bY}) = \bI_q$, and $\var(\bZ^{\text{shared}})= \bI_m$ and no mutual correlation among the components of $\bZ^{\bX}$, $\bZ^{\bY}$, and $\bZ^{\text{shared}}$. The second level latent variables are then generated by mixing shared and non-shared variables according to
\begin{equation}
\begin{split}
\tX^{\text{CCA}}_i & =  \sqrt{1 - |\lambda_i|}\, Z^{\bX}_i +
                        \sqrt{|\lambda_i|} Z^{\text{shared}}_i \\
\tY^{\text{CCA}}_i & =  \sqrt{1 -  |\lambda_i|}\ Z^{\bY}_i + 
                        \sqrt{|\lambda_i|}  Z^{\text{shared}}_i \, \sgn(\lambda_i)
\end{split} 
\label{eq:secondlevel}
\end{equation}
where the parameters $\lambda_1, \ldots, \lambda_m$ can be positive as well as negative { and range from -1 to 1}. The components $i > m$ are always non-shared and  taken from $\bZ^{\bX}$ or $\bZ^{\bY}$ as appropriate, i.e.\ as above but with
$\lambda_{i>m} = 0$.
By construction, this results in $\var(\btX^{\text{CCA}}) = \bI_p$, $\var(\btY^{\text{CCA}}) = \bI_q$ and $\cov(\tX^{\text{CCA}}_i, \tY^{\text{CCA}}_i) = \lambda_i$. Finally, the observed variables are produced by a coloring transformation  and subsequent translation
\begin{equation}
\begin{split}
\bX &= \bPhi_{\bX}^T \btX^{\text{CCA}} + \bmu_{\bX} \\
\bY &= \bPhi_{\bY}^T \btY^{\text{CCA}} + \bmu_{\bY}
\end{split}
\end{equation}

To clarify the workings behind \eqcite{eq:secondlevel} assume there are 
three uncorrelated random variables $Z_1$, $Z_2$, and $Z_3$ with mean 0 and variance 1.
We construct $X_1$ as a mixture of $Z_1$ and $Z_3$ according to 
$X_1 = \sqrt{1-\alpha} Z_1 + \sqrt{\alpha} Z_3$ where $\alpha \in [0,1]$,
and, correspondingly, $X_2$ as a mixture of $Z_2$ and $Z_3$ via
$X_2 = \sqrt{1-\alpha} Z_2 + \sqrt{\alpha} Z_3$. 
If $\alpha=0$ then $X_1=Z_1$ and $X_2=Z_2$, and if $\alpha=1$ then $X_1=X_2=Z_3$.
By design, the new variables have mean zero ($\expect(X_1) = \expect(X_2) = 0$) and unit variance ($\var(X_1) = \var(X_2)=1$). Crucially, the weight $\alpha$ of the latent variable $Z_3$ common to both mixtures induces a correlation between $X_1$ and $X_2$.  The covariance between $X_1$ and $X_2$ is $\cov(X_1, X_2) = \cov( \sqrt{\alpha} Z_3,\sqrt{\alpha} Z_3 ) = \alpha$, and since $X_1$ and $X_2$ have variance 1 we have $\cor(X_1, X_2) = \alpha$. In \eqcite{eq:secondlevel} this is further extended to allow a signed $\alpha$ and hence negative correlations.

Note that the above probabilistic model for CCA is in fact not a single model but a family of models, since we do not completely specify the underlying distributions, only their means and (co)variances. While in practice we will typically assume normally distributed generative latent variables, and hence  normally distributed observations, it is equally possible to employ other distributions for the first level latent variables. For example, a rescaled $t$-distribution with a wider tail than the normal distribution may be employed to obtain a robustified version of CCA \citep{AdroverDonato2015}.

\section*{Discussion}

\subsection*{Synthetic data}

\begin{figure}[p!]
\vskip 0.2in
\begin{center}
\centerline{\includegraphics[width=\columnwidth]{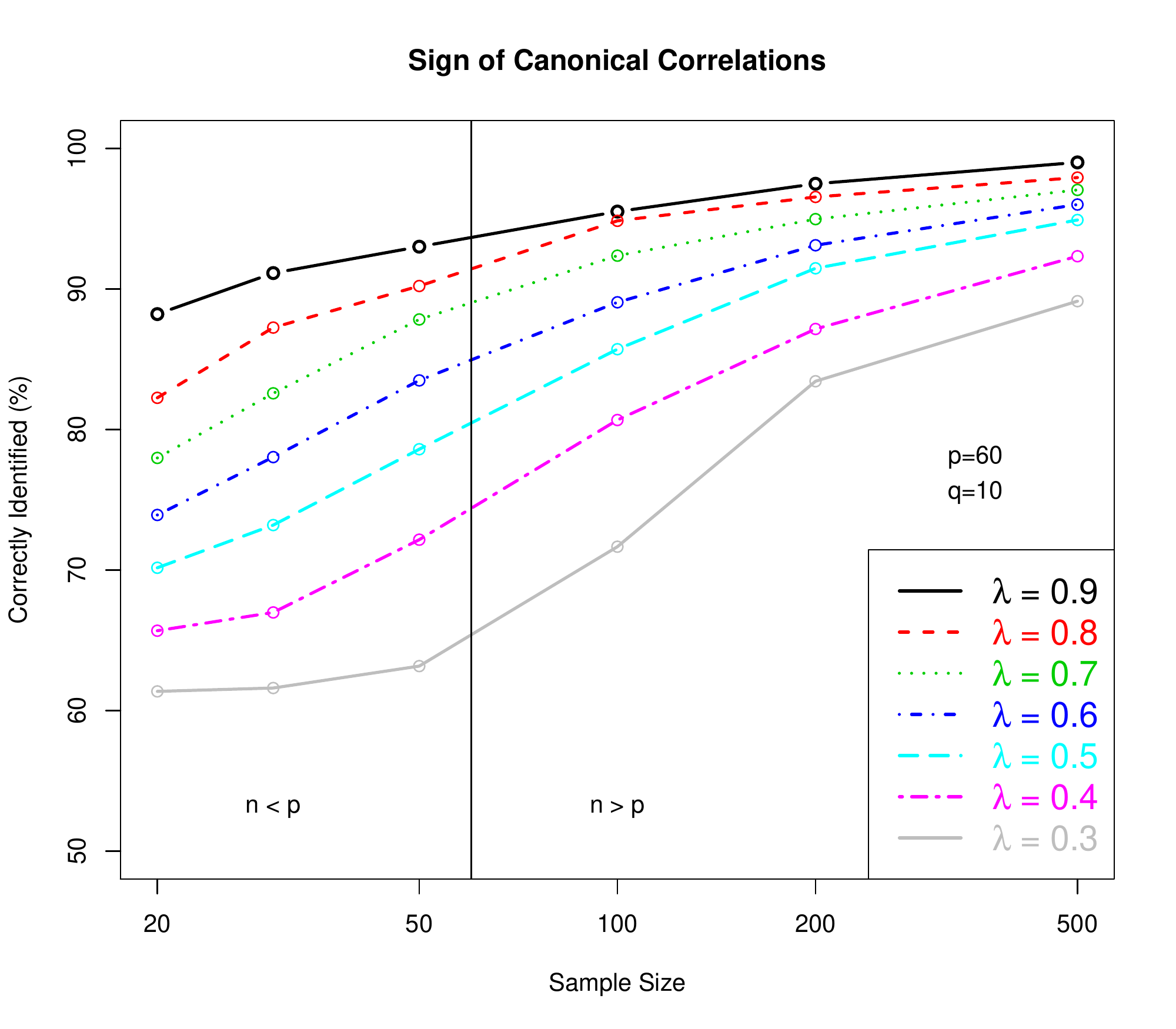}}
\caption{Percentage of estimated canonical correlations with correctly identified signs in dependence of the sample size and the strength of the true canonical correlation. \label{fig:signsim}}
\end{center}
\vskip -0.2in
\end{figure}

In order to test whether our algorithm allows to correctly identify negative canonical correlations we conducted simulations using simulated data. { Specifically,} we generated data $\bx_i$ and $\by_i$ from a $p+q$ dimensional multivariate normal distribution with zero mean and covariance matrix $\left( \begin{array}{c c} \bSigma_{\bX} & \bSigma_{\bX \bY}\\ \bSigma_{\bX \bY}^T & \bSigma_{\bY} \end{array}\right)$ where $\bSigma_{\bX} = \bI_p$, $\bSigma_{\bY} = \bI_q$ and $\bSigma_{\bX \bY} = \diag(\lambda_i)$.  The canonical correlations where set to have alternating positive and negative signs $\lambda_1=\lambda_3 =\lambda_5 =\lambda_7 = \lambda_9  = \lambda$ and $\lambda_2=\lambda_4 =\lambda_6 =\lambda_8 = \lambda_{10}  = -\lambda$ with varying strength $\lambda \in \{0.3, 0.4, 0.5, 0.6, 0.7, 0.8, 0.9\}$. 
{ A similar setup was used in \citet{CruzCanoLee2014}.} 
The dimensions were fixed at $p=60$ and $q=10$ and the sample size was $n \in\{20, 30, 50, 100, 200, 500\}$ { so that both the small and large sample regime was covered.} For each combination of $n$ and $\lambda$ the simulations were repeated 500 times, and our algorithm using shrinkage estimation of the underlying covariance matrices was employed to each of the 500 data sets to fit the CCA model.  The resulting estimated canonical correlations were then compared with the  corresponding true canonical correlations, and the proportion of correctly estimated signs was recorded.

The outcome from this simulation study is summarized graphically in \figcite{fig:signsim}. The key finding is that, depending on the strength of correlation $\lambda$ and sample size $n$, our algorithm correctly determines the sign of both negative and positive canonical correlations. As expected, the proportion of correctly classified canonical correlations increases with sample size and with the strength of correlation. Remarkably, even for comparatively weak correlation such as $\lambda=0.5$ and low sample size still the majority of canonical correction were estimated with the true sign. In short, this simulation demonstrates that if there are negative canonical correlations between pairs of canonical variables these will be detected by our approach. 

\subsection*{Nutrimouse data}

\begin{figure}[tp!]
\vskip 0.2in
\begin{center}
\centerline{\includegraphics[width=\columnwidth]{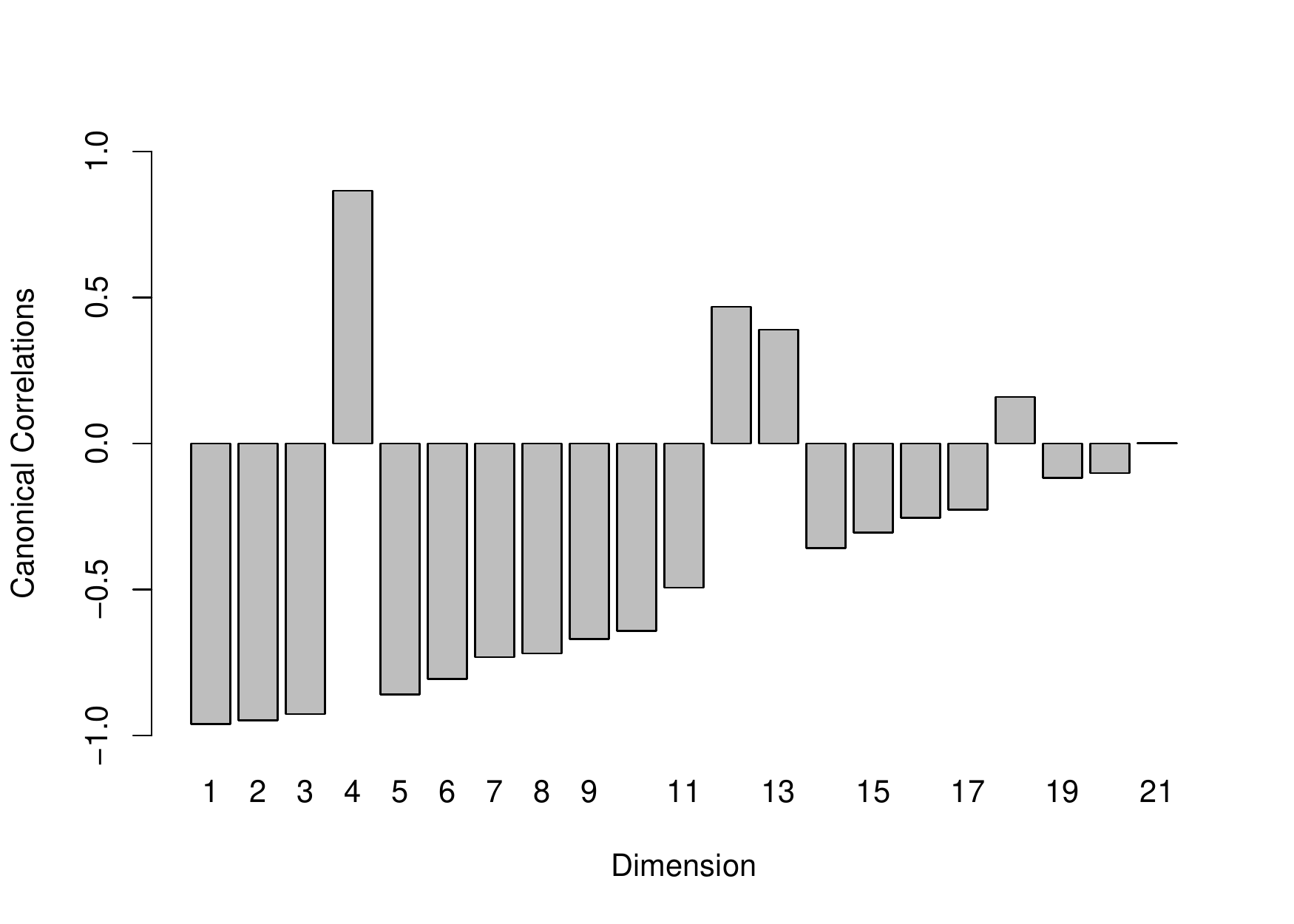}}
\caption{Plot of the estimated canonical correlations for the Nutrimouse data. The majority
of the correlations indicate a negative assocation between the corresponding canonical variables.\label{fig:nutrimousebarplot}}
\end{center}
\vskip -0.2in
\end{figure}

\begin{figure}[tp!]
\vskip 0.2in
\begin{center}
\centerline{\includegraphics[width=\columnwidth]{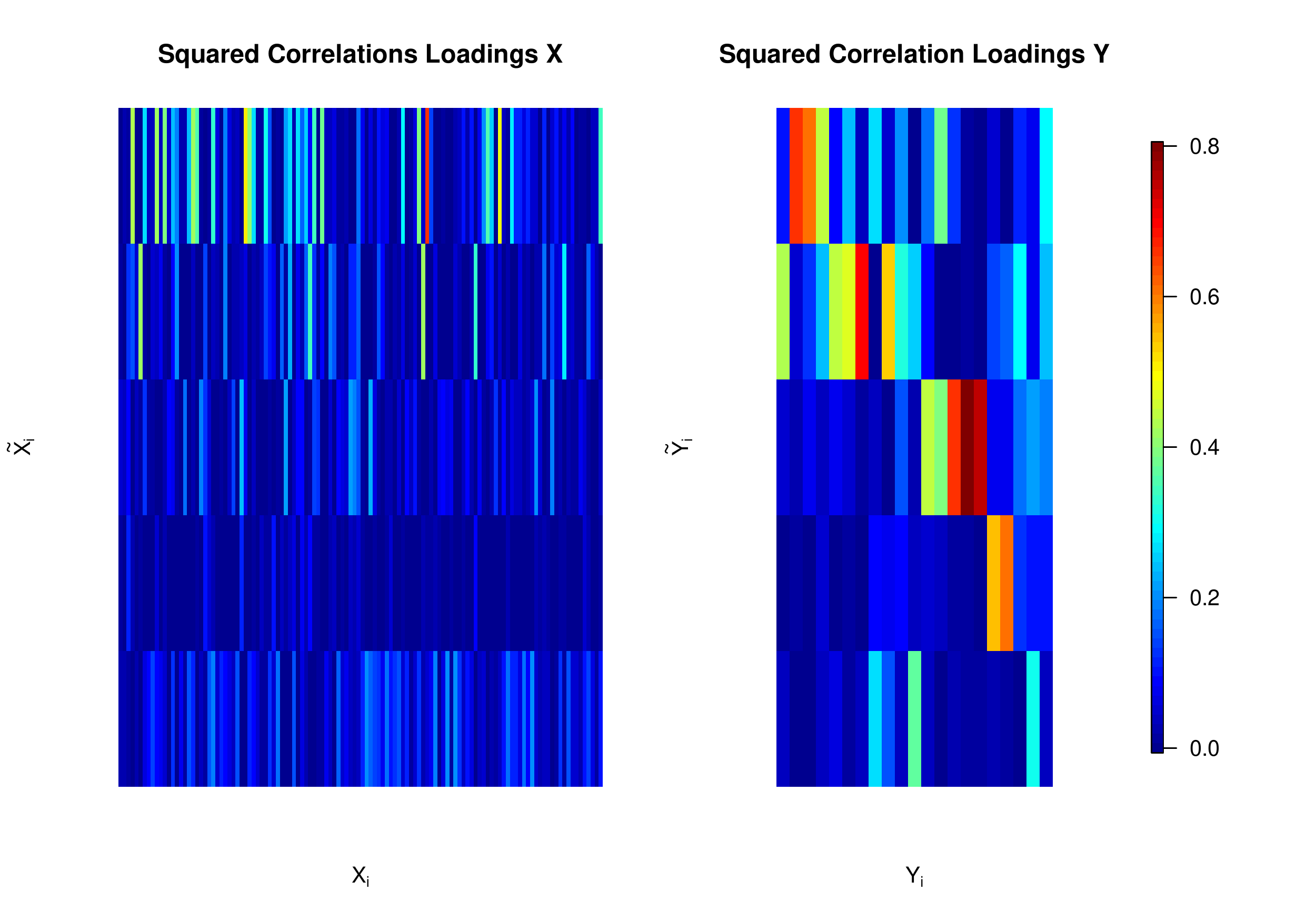}}
\caption{Squared correlations loadings between the first 5 components of the canonical covariates
$\btX^\text{CCA}$ and $\btY^\text{CCA}$ 
and the corresponding observed variables $\bX$ and $\bY$ for the Nutrimouse data.\label{fig:nutrimouseloadings}}
\end{center}
\vskip -0.2in
\end{figure}

\begin{figure}[tp!]
\vskip 0.2in
\begin{center}
\centerline{\includegraphics[width=\columnwidth]{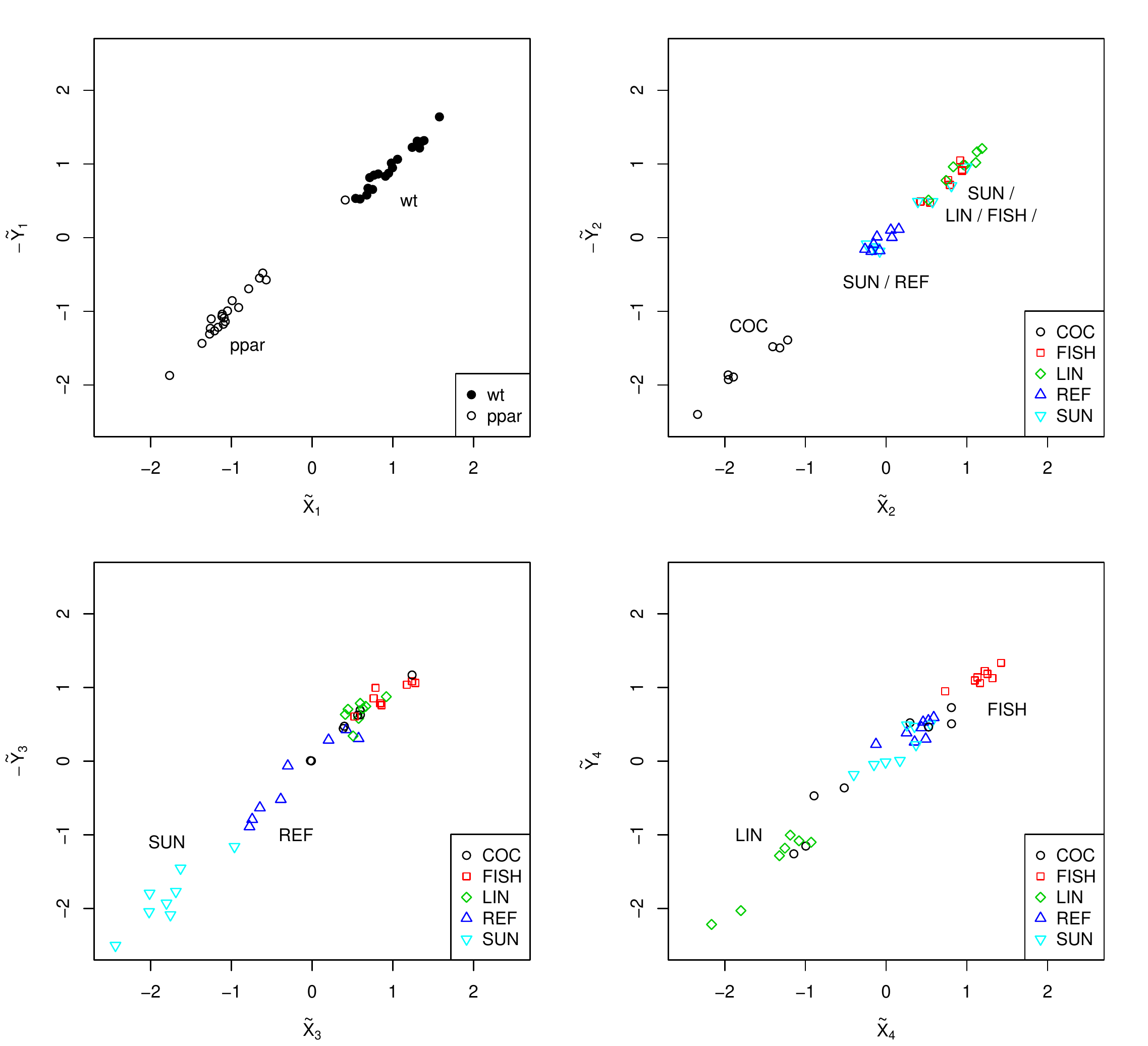}}
\caption{Scatter plots between corresponding pairs of canonical covariates for the Nutrimouse data.\label{fig:nutrimousebetween}}
\end{center}
\vskip -0.2in
\end{figure}

\begin{figure}[tp!]
\vskip 0.2in
\begin{center}
\centerline{\includegraphics[width=\columnwidth]{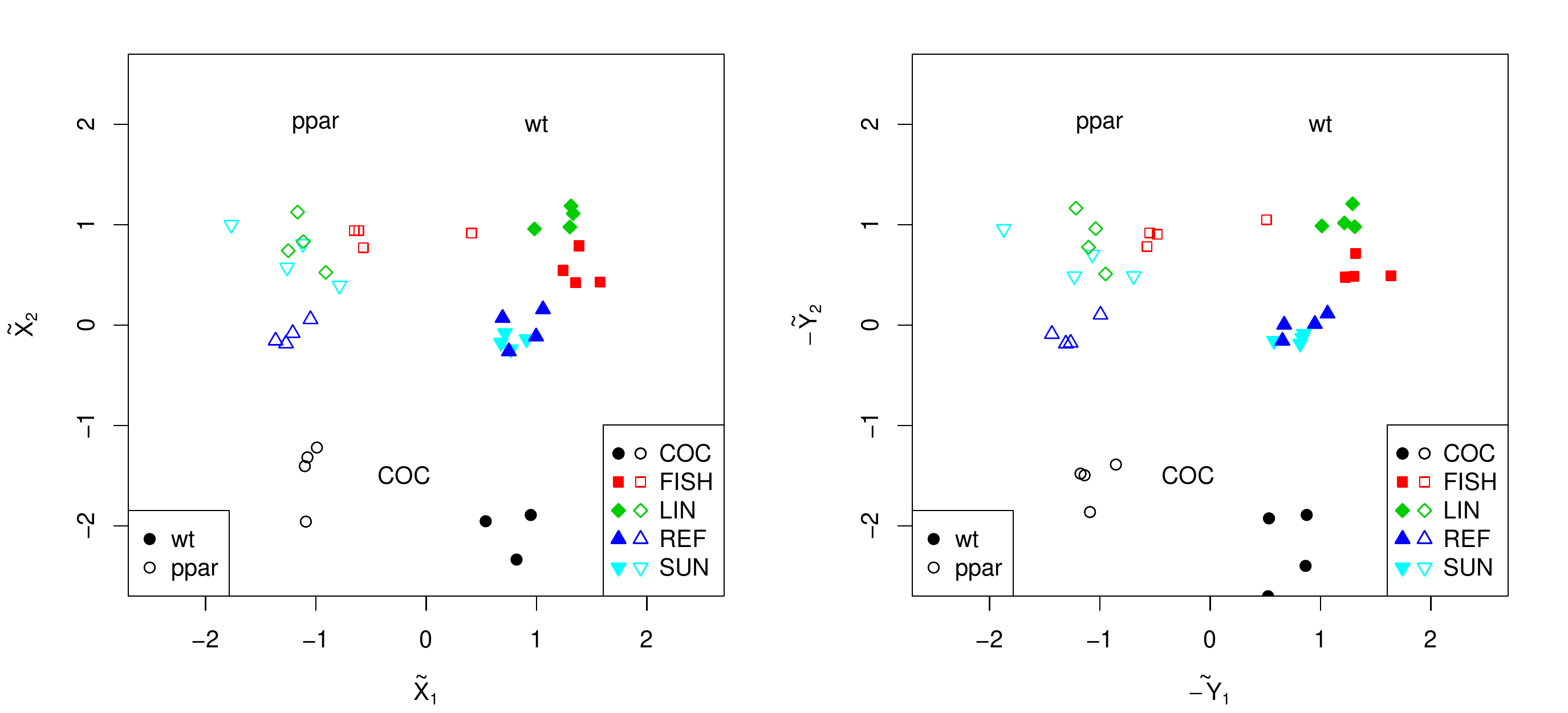}}
\caption{Scatter plots between first and second components within each
canonical covariate for the Nutrimouse data.\label{fig:nutrimousewithin}}
\end{center}
\vskip -0.2in
\end{figure}

{ We now analyze two experimental omics data sets to illustrate our approach. Specifically, we demonstrate the capability of our variant of CCA to identify negative canonical correlations among canonical variates as well its application to high-dimensional data where the number of samples $n$ is smaller than the number of variables $p$ and $q$.}

The first data set is due to \citet{Martin+2007} and results from a nutrigenomic study in the mouse studying $n=40$ animals.  The $\bX$ variable collects the measurements of the gene expression of $p=120$ genes in liver cells. These were selected a priori considering the biological relevance for the study. The $\bY$ variable contains lipid concentrations of $q=21$ hepatic fatty acids, measured on the same animals. Before further analysis we standardized both $\bX$ and $\bY$.

Since the number of available samples $n$ is smaller than the number of genes $p$ we used shrinkage estimation to obtain the joint correlation matrix which resulted in a shrinkage intensity of $\lambda_{\text{cor}}=0.16$.     Subsequently, we computed canonical directions and associated canonical correlations $\lambda_1,\dots, \lambda_{21}$.  The canonical correlations are shown in \figcite{fig:nutrimousebarplot}, and range in value between -0.96 and 0.87.  As can be seen, 16 of the 21 canonical correlations are negative, including the first three top ranking correlations. In \figcite{fig:nutrimouseloadings} we depict the squared correlation loadings between the first 5 components of the canonical covariates $\btX^\text{CCA}$ and $\btY^\text{CCA}$ and the corresponding observed variables $\bX$ and $\bY$. This visualization shows that most information about the correlation structure within and between the two data sets (gene expression and lipid concentrations) is concentrated in the first few latent components.

This is confirmed by further investigation of the scatter plots both between corresponding pairs of $\btX^\text{CCA}$ and $\btY^\text{CCA}$ canonical variates (\figcite{fig:nutrimousebetween}) as well as within each variate  (\figcite{fig:nutrimousewithin}).  Specifically, the first CCA component allow to identify the genotype of the mice (wt: wild type; ppar: PPAR-$\alpha$ deficient)  whereas the subsequent few components  reveal the imprint of the effect of the various diets (COC: coconut oil; FISH: fish oils; LIN: linseed oils; REF: reference diet; SUN: sunflower oil)  on gene expression and lipid concentrations.  
%Since the separation in particular of the genotypes is strong  we suspect that in this data set one FISH eating wild type mouse may have been wrongly classified as a PPAR-$\alpha$ deficient mouse (this corresponds to the ppar data point in the wt cluster in \figcite{fig:nutrimousebetween}~top-left and in \figcite{fig:nutrimousewithin}).

%\clearpage

\subsection*{The Cancer Genome Atlas LUSC data}

As a further illustrative example we studied genomic data from The Cancer Genome Atlas (TCGA), a public resource that catalogues clinical data and molecular characterizations of many cancer types \citep{Kandoth+2013}.  We used the TCGA2STAT tool to access the TCGA database from within R \citep{WanAllenLiu2016}

Specifically, we retrieved gene expression (RNASeq2) and methylation data for lung squamous cell carcinoma (LUSC) which is one of the most common types of lung cancer. After download, calibration and filtering as well as matching the two data types to 130 common patients following the guidelines in \citet{WanAllenLiu2016} we obtained two data matrices, one ($\bX$) measuring gene expression  of $p=206$ genes and one ($\bY$) containing methylation levels corresponding to $q=234$ probes. As clinical covariates the sex of each of the 130 patients {(97 males, 33 females)} was downloaded as well as the vital status {(46 events in males, and 11 in females)} and cancer end points, i.e. the number of days to last follow-up or the days to death.  In addition, since smoking cigarettes is a key risk factor for lung cancer, the number of packs per year smoked was also recorded.
{ The number of packs ranged from 7 to 240, so all of the patients for
which this information was available were smokers.}

\begin{figure}[tp!]
\vskip 0.2in
\begin{center}
\centerline{\includegraphics[width=\columnwidth]{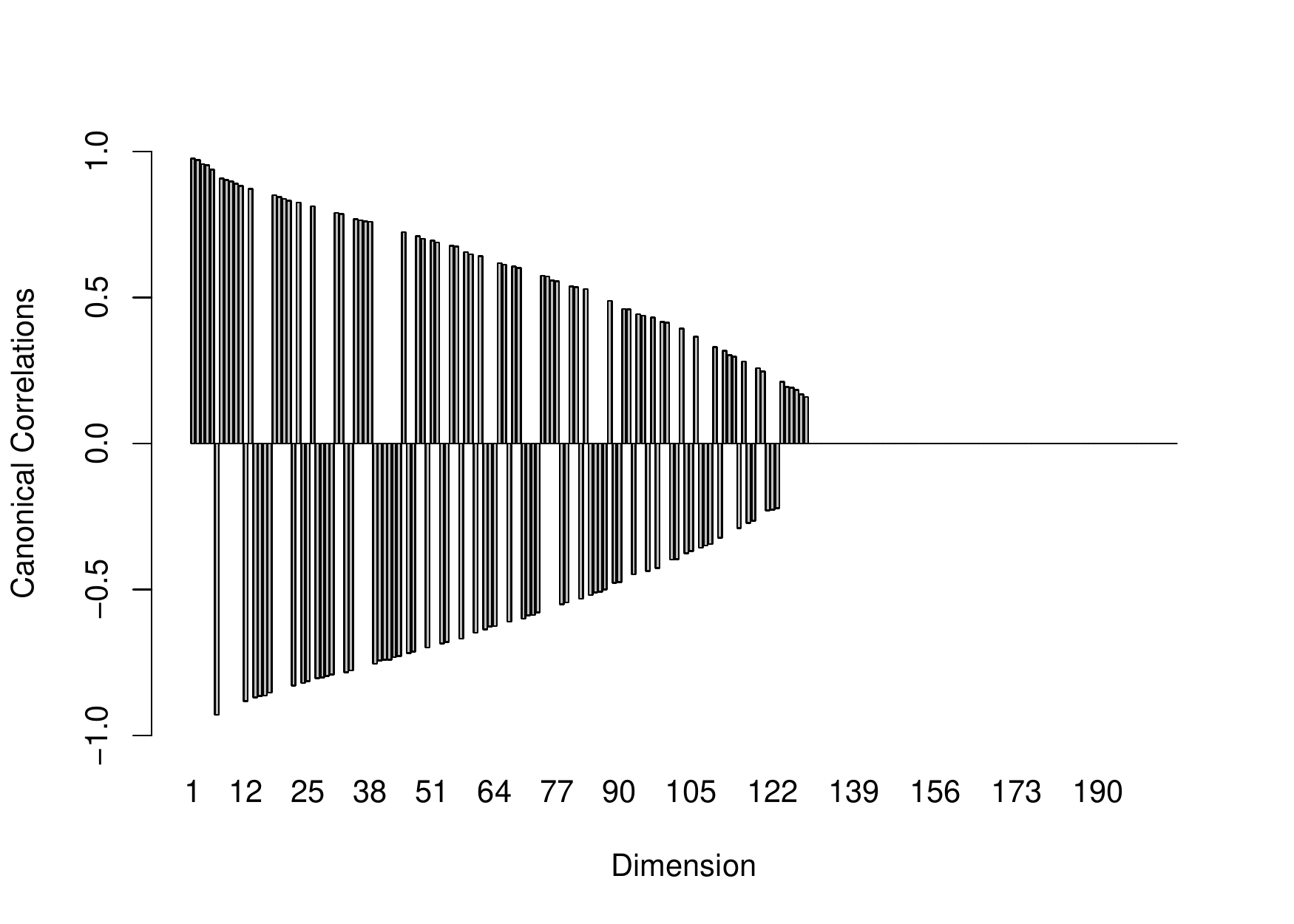}}
\caption{Plot of the estimated canonical correlations for the TCGA LUSC data.\label{fig:luscbarplot}}
\end{center}
\vskip -0.2in
\end{figure}

\begin{figure}[tp!]
\vskip 0.2in
\begin{center}
\centerline{\includegraphics[width=\columnwidth]{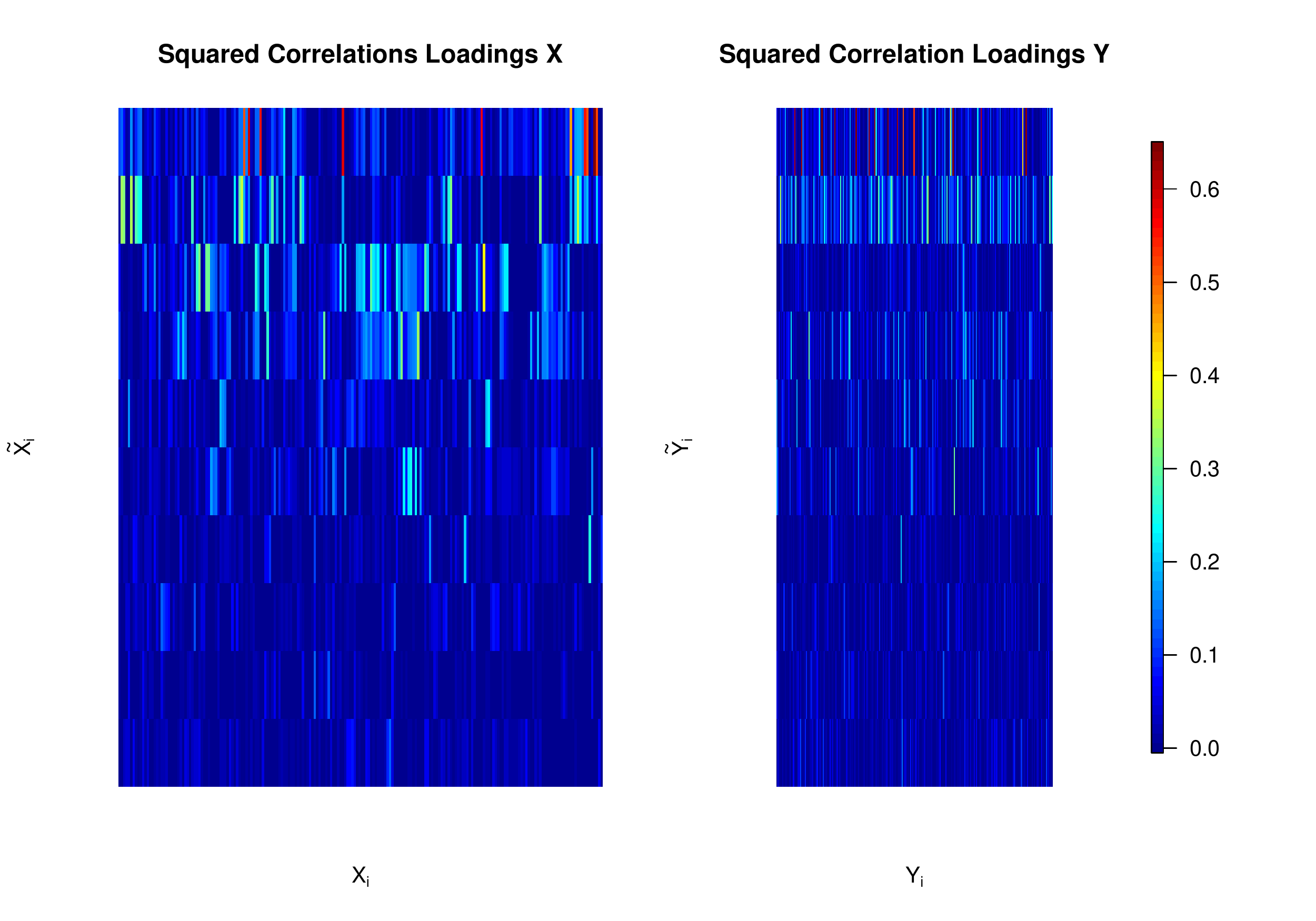}}
\caption{Squared correlations loadings between the first 10 components of the canonical covariates
$\btX^\text{CCA}$ and $\btY^\text{CCA}$ 
and the corresponding observed variables $\bX$ and $\bY$ for the TCGA LUSC data.\label{fig:luscloadings}}
\end{center}
\vskip -0.2in
\end{figure}

\begin{figure}[tp!]
\vskip 0.2in
\begin{center}
\centerline{\includegraphics[width=\columnwidth]{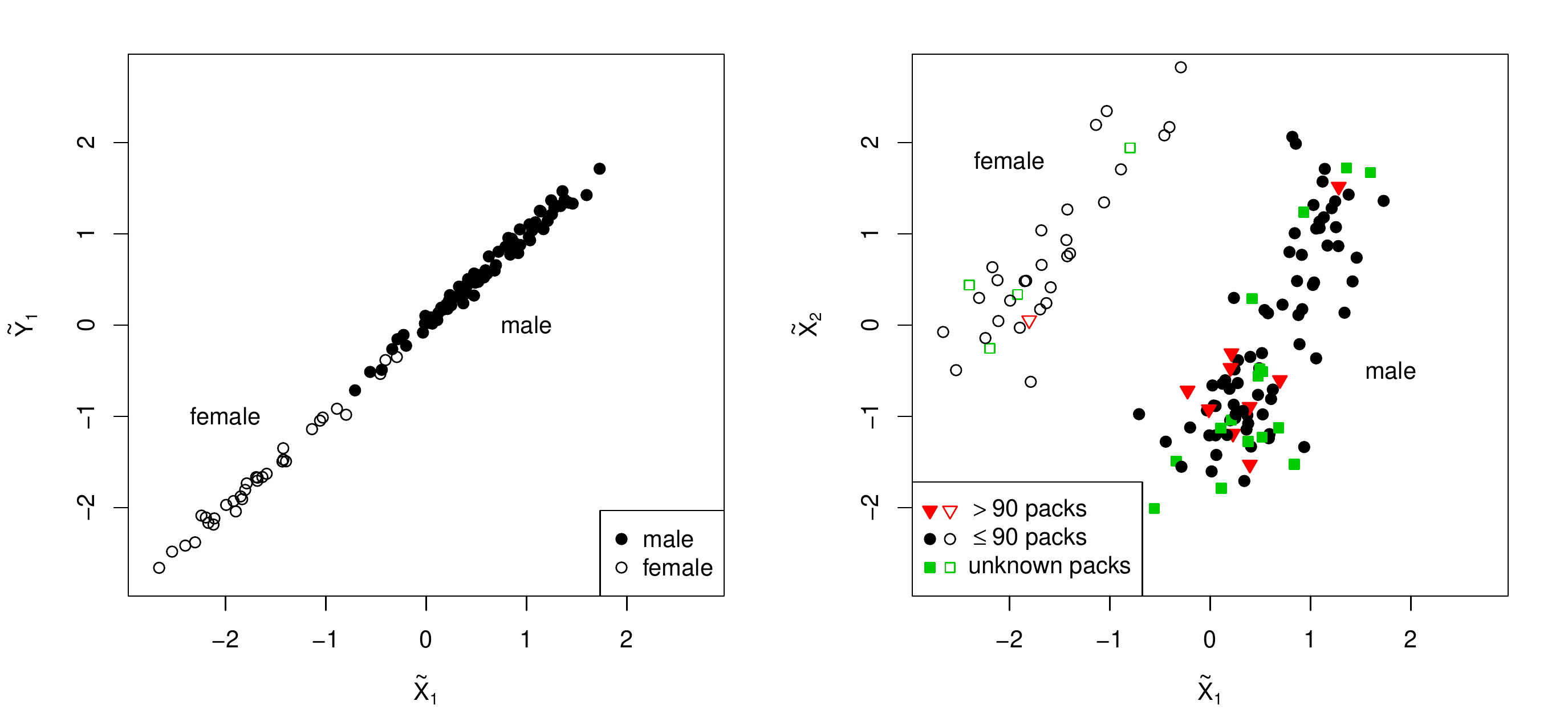}}
\caption{Scatter plots between first component of $\btX^\text{CCA}$ and $\btY^\text{CCA}$ (left)
 and within the first two components of $\btX^\text{CCA}$ (right) for the TCGA LUSC data.\label{fig:luscscatter}}
\end{center}
\vskip -0.2in
\end{figure}

\begin{figure}[tp!]
\vskip 0.2in
\begin{center}
\centerline{\includegraphics[width=\columnwidth]{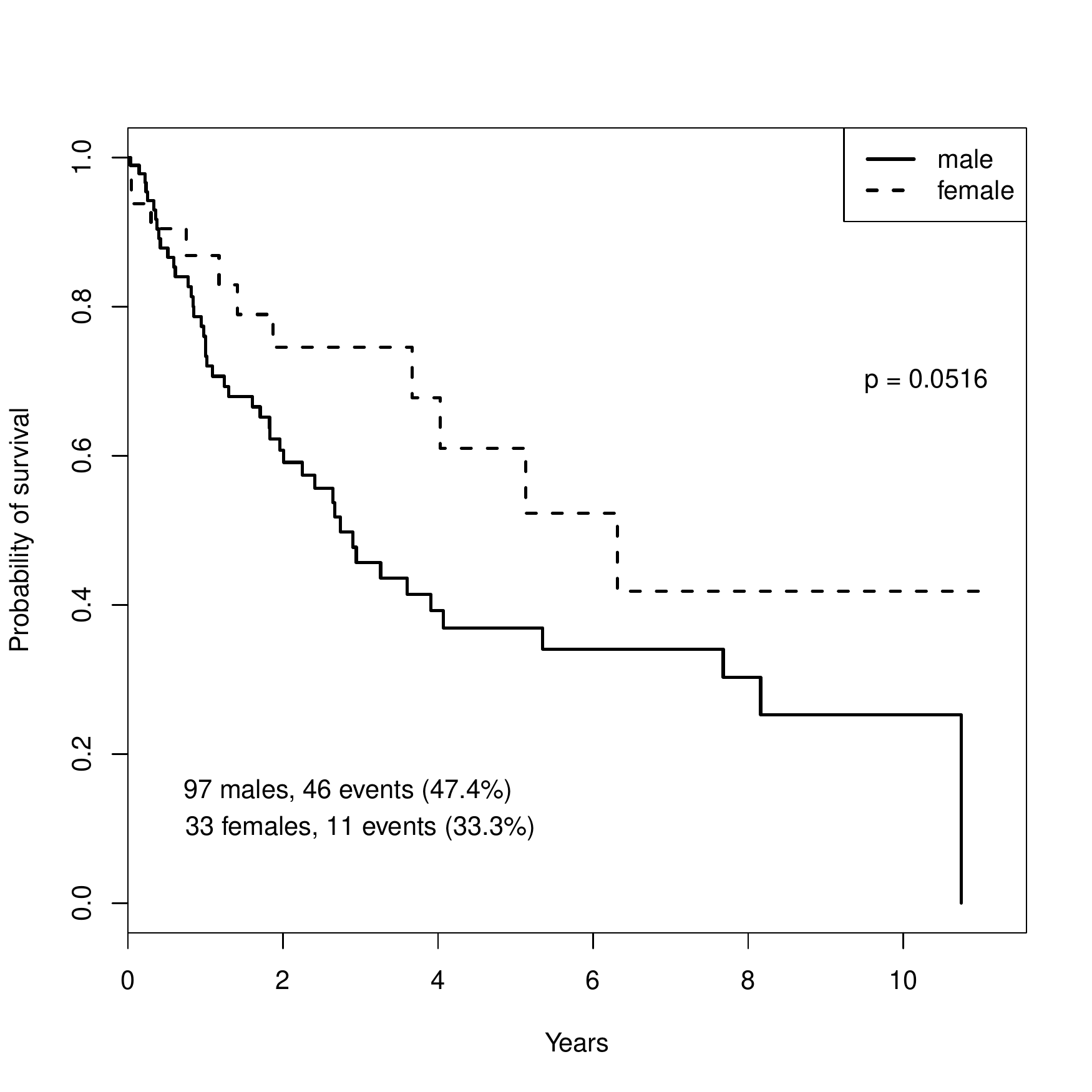}}
\caption{Plot of the survival probabilities for male and female patients for the TCGA LUSC data.\label{fig:luscsurv}}
\end{center}
\vskip -0.2in
\end{figure}

As above we applied the shrinkage CCA approach to the LUSC data which resulted in a correlation shrinkage intensity of $\lambda_{\text{cor}}=0.19$.   Subsequently, we computed canonical directions and associated canonical correlations $\lambda_1,\dots, \lambda_{21}$. The canonical correlations are shown in \figcite{fig:luscbarplot}, and range in value between -0.92 and 0.98. Among the top 10 strongest correlated pairs of canonical covariates only one has a negative coefficient.  The plot of the squared correlation loadings  (\figcite{fig:luscloadings}) for these 10 components already indicates that the data can be sufficiently summarized by a few canonical covariates.

Scatter plots between the first pair of canonical components and between the first two components of $\btX^\text{CCA}$ are presented in \figcite{fig:luscscatter}.  These plots show that the first canonical component corresponds to the sex of the patients, with males and females being clearly separated by underlying patterns in gene expression and methylation. The survival probabilities computed for both groups show that there is a statistically significant different risk pattern between males and females (\figcite{fig:luscsurv}).  However, inspection of the second order canonical variates reveals  that the difference in risk is likely due to overrepresentation of strong smokers in male patients rather than being directly attributed to the sex of the patient (\figcite{fig:luscscatter}~right).

%\clearpage

\section*{Conclusions}

CCA is crucially important procedure for integration of multivariate data. Here, we have revisited CCA from the perspective of whitening that allows a better understanding of both classical CCA and its probabilistic variant. In particular, our main contributions in this paper are:
\begin{itemize}
\item first, we show that CCA is procedurally equivalent to a special whitening transformation, that unlike other general whitening procedures, is uniquely  defined and without any rotational ambiguity;
\item second, we demonstrate the direct connection of CCA with multivariate regression and demonstrate that CCA is effectively a linear model between whitened variables, and that correspondingly canonical correlations are best understood as regression coefficients;
\item third, the regression perspective advocates for permitting both positive and negative canonical correlations and we show that this also allows to resolve the sign ambiguity present in the canonical directions;
\item fourth, we propose an easily interpretable probabilistic generative model for CCA as a two-layer latent variable framework that not only admits canonical correlations of both signs but also allows non-normal latent variables; 
\item and fifth, we provide a computationally effective computer implementation in the ``whitening'' R package based on high-dimensional shrinkage estimation of the underlying covariance and correlation matrices and show that this approach performs well both for simulated data as well as in application to the analysis of various types of omics data.
\end{itemize}
In short, this work provides a  unifying perspective on CCA, linking together sphering procedures, multivariate regression and corresponding  probabilistic generative models, and also offers a practical tool for high-dimensional CCA for practitioners in applied statistical data analysis.

\section*{Methods}

\subsection*{Implementation in R}

We have implemented our method for high-dimensional CCA allowing for potential negative canonical correlations in the R package ``whitening'' that is freely available from \url{https://CRAN.R-project.org/package=whitening}.  The functions provided in this package incorporate the computational efficiencies described below. The R package also includes example scripts.  The ``whitening'' package has been used to conduct the data analysis described in this paper. Further information and R code to reproduce the analyses in this paper is available at \url{http://strimmerlab.org/software/whitening/}.

\subsection*{High-dimensional estimation}

Practical application of CCA, in both the classical and probabilistic variants, requires estimation of the joint covariance of $\bX$ and $\bY$ from data, as well as the computation of the corresponding underlying whitening matrices $\bW_{\bX}^\text{CCA}$ and $\bW_{\bY}^\text{CCA}$ (i.e. canonical directions) and canonical correlations $\lambda_i$.

In moderate dimensions and large sample size $n$, i.e.\ when both $p$ and $q$ are not excessively big and $n$ is larger than both $p$ and $q$ the classic CCA algorithm is applicable and empirical or maximum likelhood estimates may be used.   Conversely, if the sample size $n$ is small compared to $p$ and $q$ then there exist numerous effective Bayesian, penalized likelihood and other related regularized estimators to obtain \emph{statistically efficient estimates} of the required covariance matrices \citep[e.g.][]{SS05C,BickelLevina2008,HannartNaveau2014,Touloumis2015}.  In our implementation in R and in the analysis below we use the shrinkage covariance estimation approach developed in \citet{SS05C} and also employed for CCA analysis in \citet{CruzCanoLee2014}.
{ However, in principle any other preferred covariance estimator may be applied.}

\subsection*{Algorithmic efficiencies}

In addition to statistical issues concerning accurate estimation,  high dimensionality  also poses substantial challenges in \emph{algorithmic} terms, with regard both to memory requirements as well as to computing time.  Specifically, for large values of $p$ and $q$ directly performing the matrix operations necessary for CCA, such as computing the matrix square root or even simple matrix multiplication, will be prohibitive since these procedures typically scale in cubic order of $p$ and $q$.

In particular, in a CCA analysis this affects  i) the computation and estimation of the matrix $\bK$ (\eqcite{eq:Kdef}) containing the adjusted cross-correlations, and ii) the calculation of the whitening matrices $\bW_{\bX}^\text{CCA}$ and $\bW_{\bY}^\text{CCA}$ with the canonical directions $\balpha_i$ and $\bbeta_i$ from the rotation matrices $\bQ_{\bX}^\text{CCA}$ and  $\bQ_{\bY}^\text{CCA}$ (\eqcite{eq:whitencor}).  These computational steps involve  multiplication and square-root calculations involving possibly very large matrices of dimension $p\times p$ and $q\times q$.

Fortunately, in the small sample domain with $n \leq p,q$ there exist computational tricks to perform these matrix operations in a very effective and both time- and memory-saving manner that avoids to directly compute and handle the large-scale covariance matrices and their derived quantities \citep[e.g.][]{HT04}.  Note this requires the use of regularized estimators, e.g. shrinkage or ridge-type estimation.  Specifically, in our implementation of CCA we capitalize on an algorithm described in \citet{Zuber+2012} (see next section for details) that allows to compute the matrix product of the inverse matrix square root of the shrinkage estimate of the correlation matrix $\bR$ with a matrix $\bM$ without the need to store or compute the full estimated correlation matrices.  The computationals savings due to effective matrix operations for $n < p$ and $n < q$ can be substantial, going from $O(p^3)$ and $O(q^3)$ down to $O(n^3)$ in terms of algorithmic complexity. Correspondingly, for example for $p/n$ = 3 this implies time savings of factor 27 compared to ``naive'' direct computation.

\subsection*{Zuber et al. algorithm}

\citet{Zuber+2012} describe an algorithm that allows to compute the matrix product of the inverse matrix square root of the shrinkage estimate of the correlation matrix $\bR$ with a matrix $\bM$ without the need to store or compute the full estimated correlation matrices. Specifically, writing the correlation estimator in the form
\begin{equation}
\underbrace{\bR}_{p \times p} =  \lambda (\bI_p + \underbrace{\bU}_{p \times n} \underbrace{\bN}_{n \times n} \bU^T ) 
\end{equation}
allows for algorithmically effective matrix multiplication of
\begin{equation}
\underbrace{\bR^{-1/2}}_{p \times p} \underbrace{\bM}_{p \times d} = 
\lambda^{-1/2} ( \bM - \bU\underbrace{ (\bI_n-(\bI_n + \bN)^{-1/2}}_{n \times n} ) (\underbrace{\bU^T  \bM}_{n \times d} )) \,.
\end{equation}
Note that on the right-hand side of these two equations no matrix of dimension $p \times p$ appears; instead all matrices are of much smaller size. 

In the CCA context we apply this procedure to \eqcite{eq:whitencor} in order to obtain the whitening matrix and the canonical directions and also to \eqcite{eq:Kdef} to efficiently compute the matrix $\bK$.

\section*{Acknowledgements}
KS thanks Martin Lotz for discussion.
TJ was funded by a Wellcome Trust ISSF Ph.D.\ studentship.
We also thank the anonymous referees for their many useful suggestions.

%\newpage

\bibliographystyle{apalike}
\bibliography{preamble,stats,strimmer,omics}

\end{document}